\documentclass[letterpaper,twocolumn,10pt]{article}
\usepackage{usenix}

\usepackage{tikz}
\usepackage{amsmath}

\usepackage{graphicx}
\usepackage{multirow}
\usepackage{algorithm}
\usepackage{algpseudocode}
\usepackage{booktabs}

\usepackage{filecontents}

\begin{document}

\date{}

\title{\Large \bf Trident: Improving Malware Detection with LLMs and Behavioral Features}

 \author{
 {\rm Rebecca Saul}\\
 University of California, Berkeley
 \and
 {\rm Jingzhi Jiang}\\
 University of California, Berkeley
 \and
 {\rm Elliott Chia}\\
 University of California, Berkeley
 \and
 {\rm David Wagner}\\
 University of California, Berkeley
}

\maketitle

\begin{abstract}
Traditionally, machine learning methods for PE malware detection have relied on static features like byte histograms, string information, and PE header contents. One barrier to incorporating dynamic analysis features has been the semi-structured nature of sandbox behavior reports. We show that, using the latest generation of large language models with reasoning, it is possible to efficiently process these behavior reports and utilize them as part of a malware detection pipeline. Specifically, we leverage LLMs to generate behavior-based malware detection rules based on a small training set of labeled malware. We find that these detection rules, derived from behavioral features, are much more robust to concept drift than standard static-feature methods, while maintaining practical false positive rates. Finally, we introduce Trident, a system which combines a classic decision tree model over static features, our behavior-based detection rules, and direct LLM analysis of sandbox reports through majority voting. Trident outperforms standard methods using static features, outperforms behavior-based rules alone, and is as resilient to concept drift as active learning methods without requiring retraining.
\end{abstract}

\section{Introduction}

\begin{table}
\small
    \centering
    \begin{tabular}{l c c c}
        \toprule
        \textbf{Method} & \textbf{Recall} & \textbf{F1} & \textbf{FPR} \\ 
        \midrule
        Behavioral rules alone (ours)   & 0.942 & 0.964 & 0.0160 \\ 
        
        Neurlux (ML + behavior features) & 0.925 & 0.957 & 0.0097 \\
        Bag-of-Behaviors + LR & 0.890 & 0.928 & 0.0319 \\
        Bag-of-Behaviors + GBDT & 0.913 & 0.945 & 0.0251 \\
        
        Behavioral rules+LLM (ours)     & 0.936 & 0.966 & 0.0024 \\ 
        ML classifier, static feats. (GBDT) & 0.936 & 0.966 & 0.0010 \\ 
        GBDT with monthly updates & 0.958 & 0.978 & 0.0008 \\ 
        Trident (ours)      & \textbf{0.964} & \textbf{0.982} & \textbf{0.0003} \\ 
        
         \bottomrule
    \end{tabular}
    \caption{Performance of eight malware detectors on BODMAS, averaged over all months.}
    \label{tab:results-summary}
\end{table}

\begin{figure*}[h]
    \centering
    \includegraphics[width=\linewidth]{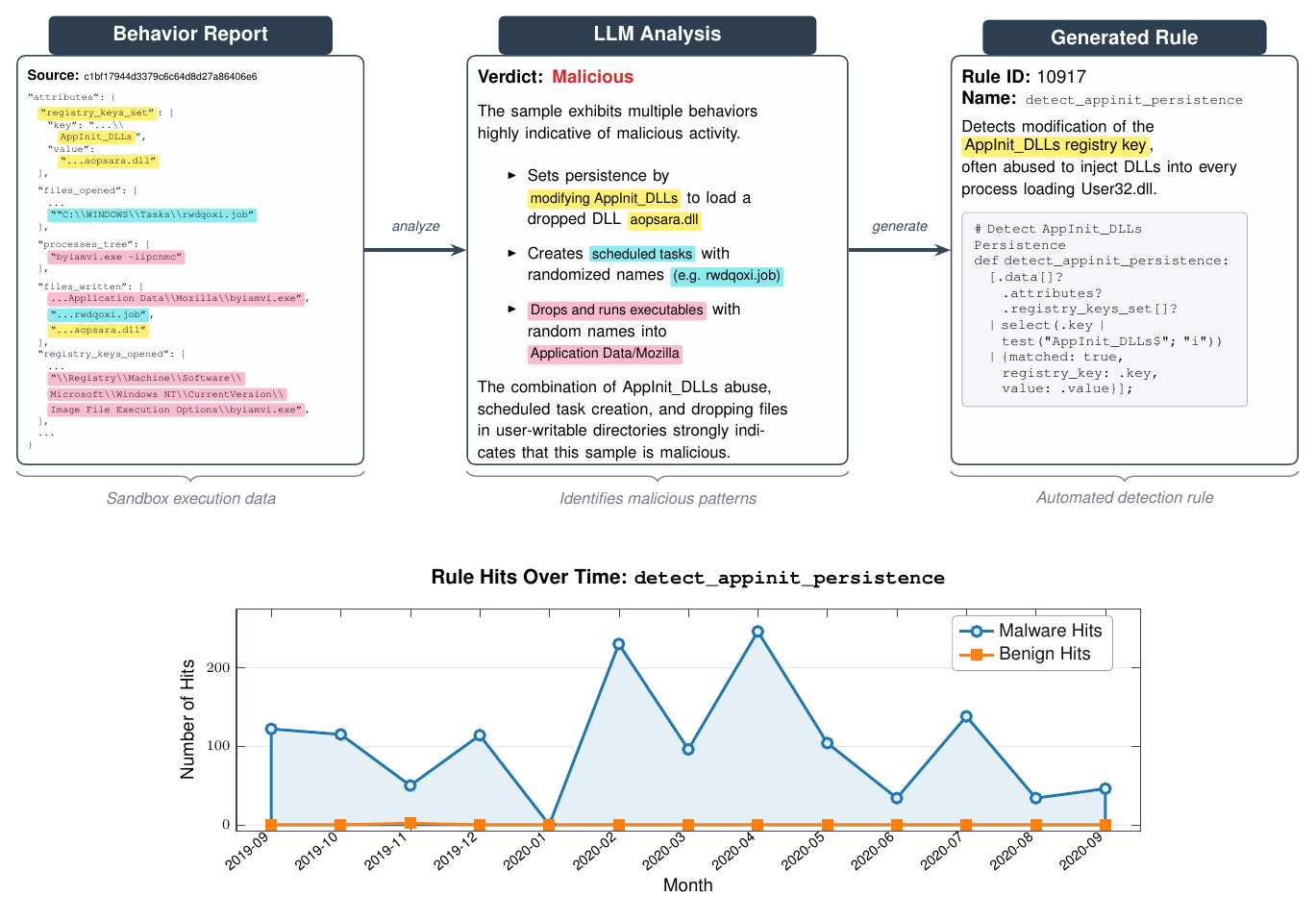}
    \caption{The LLM analyzes a malware sample's behavior report (panel 1) and identifies suspicious behaviors (panel 2). It then formalizes these behaviors into malware detection rules (panel 3). These rules generalize to new malware samples while maintaining low false positive rates (bottom panel).}
    \label{fig:mainfig}
\end{figure*}

Malware detection and analysis is an enduring problem in cybersecurity. As computer systems and the internet become more integral to daily life, the threat posed by malicious software, and the volume of attacks, has increased. In 2025, Microsoft reported blocking 4.5 million new malware files \emph{every day}~\cite{microsoft_malware_report}, and the United States Federal Bureau of Investigation attributed over \$50M in losses to malware and ransomware~\cite{fbi_report_cybercrime}. Moreover, as adversaries gain access to AI tools with the capacity to automate attacks, the scale and sophistication of malware campaigns is projected to increase. In this paper, we show that advances in AI can also benefit defenders. Specifically, we demonstrate how modern LLMs enable a new approach to malware detection that is fully automated, high performing, interpretable, and addresses the problem of concept drift.

One simple approach to malware detection uses static signatures: collect samples of known malware, ask human analysts to identify a short byte sequence (the signature) that is unique to each sample, and block any binary that contains this signature~\cite{signatures1,signatures2}.
Unfortunately, static signatures are easy to evade, not effective at detecting new malware, and often require substantial human effort to create.

The security community has explored many alternatives to address these shortcomings.

In the past decade, the collection and publication of large, open source, malware datasets has facilitated research into a variety of different machine learning (ML) models for malware detection~\cite{ember,sorel,bodmas}. ML approaches have enabled malware detection to scale to the big data era, but several challenges remain:

\begin{enumerate}
    \item \emph{Reliance on static features}. Many models rely on features extracted during static analysis of an executable. While cheap to retrieve, these features are often brittle and more easily manipulated by malware authors. This lowers the barrier for adversaries looking to evade classifiers reliant on such features.

    \item \emph{Vulnerability to concept drift}. Unlike with traditional classification tasks, the malware data landscape is rapidly and continually evolving as attackers try to outwit the latest detection models and defenders attempt to keep pace. While models often achieve extremely high accuracy on their validation sets, once deployed, performance tends to decline over time as the underlying data distribution shifts. 

    \item \emph{Expensive upkeep costs}. Due to concept drift, to maintain acceptable detection and false positive rates, models need to be regularly retrained with new data. Provisioners can accrue significant expenses while collecting and labeling new data, in addition to the model training/compute itself.

    \item \emph{Lack of interpretability}. The neural architectures used for malware classification face the same interpretability difficulties as those used in other machine learning settings. Thus, when a model makes a classification decision, there is little explanation available to analysts as to what features the model relied on to come to that conclusion. This limits the insights that analysts can gain from using these models and makes it hard for analysts to audit model outputs.

\end{enumerate}

To address these challenges, we propose a new malware detection framework that leverages large language models (LLMs) to write malware detection rules and make classification decisions. We characterize files of interest using behavior reports generated from running the samples in a sandbox. Our method consists of two phases. In Phase I (Initialization), we utilize a frontier LLM to analyze the sandbox report of a malware sample, identify the malicious behaviors it exhibits, and craft rules that detect those behaviors (Figure~\ref{fig:mainfig}). In this manner, we curate an initial set of malware detection rules from an expert-labeled dataset. In Phase II (Deployment), we execute these rules against unknown samples to get malicious/benign labels. On files where the rules do not give a definitive verdict, we ask an LLM for a second opinion.

One technical complication with LLM-generated rules is that the quality of rules created in this way can vary.
Some LLM-written rules have very high false positive rates, so naive use of a LLM yields a poor solution.
We develop a verification procedure to validate LLM-written rules, discarding any that trigger false positives on benign training samples.

We show that these methods can achieve low false positive rates ($\sim$ 0.2\%), close to that achievable with static ML classifiers. Moreover, these behavioral rules are interpretable and capture the actions that were malicious or suspicious themselves---a stark contrast from static features, like high entropy, that are often correlated with malware but not in themselves problematic. 

Compared to classifying based on static features, we find that our method is more often able to detect new malware and new malware families. This occurs because new malware often shares some malicious behaviors with older malware and is thus detected by rules crafted from past known malware samples. In this way, the use of dynamic features translates into improved resilience to concept drift, without the need for frequent, costly, model retraining. 

Finally, we demonstrate that the best performance is achieved by combining LLM-generated behavioral rules and a standard static ML classifier.
Our system, Trident, performs better on the BODMAS dataset than a static classifier ($3\times$ lower FPR, $1.7\times$ lower FNR), and better than behavioral rules alone (Table~\ref{tab:results-summary}).
Trident also resolves concept drift problems that plague ML classifiers on BODMAS, maintaining steady performance over time, while a static ML classifier suffers a significant drop in performance after one year (Trident F1: $0.975 \to 0.975$, static ML F1: $0.975 \to 0.91$; Figure~\ref{fig:trident_results}).

Our work suggests that LLMs enable new methods for malware detection and that behavioral features can mitigate a critical vulnerability, concept drift, of current ML detection systems.

\section{Background}
\subsection{Malware Detection from Static Features} 

Static malware detection classifies samples as benign or malicious without executing them. The classifier uses static features, including PE headers, strings, imported functions, and byte sequences~\cite{ember}, to find anomalies in samples. Although static features are more easily manipulated by attackers, static detection has the major benefit of identifying malicious files without running them, which is time-consuming and requires intricate sandbox setups to do safely. 

One of the main static malware detection methods is machine learning-based malware detection. Static PE malware detectors appeared at least as early as 2001, when Schultz et al.~\cite{schultz_data_2001} used static features (PE executable text headers and byte sequences) to train data mining models like RIPPER~\cite{cohen_learning_1996} and Naive Bayes. Subsequent research has followed similar logic, training various machine learning models on more updated and diverse malware datasets. 

In 2017, Anderson and Roth~\cite{ember} released the EMBER dataset, which contains benign and malicious files from sources like VirusTotal~\cite{virustotal}, VX Heaven, and VirusShare~\cite{noauthor_virussharecom_nodate}. EMBER also defined eight groups of features that can be extracted from each Windows PE binary: general file information, header information, imported functions, exported functions, section information, raw byte histogram, byte entropy histogram, and string extraction. The authors train a gradient-boosted decision tree (GBDT) on these features.
Several year later, researchers released the BODMAS~\cite{bodmas} dataset with new malware and benign samples and malware family labels.

\subsection{Malware Detection from Dynamic Features} 
While static analysis is a powerful tool, there are a number of methods attackers can use to defeat static detection. Code obfuscation, encryption, and packing let malware developers alter the static features of their samples without modifying their underlying behavior~\cite{obfuscation,packing}. However, it is much more difficult for attackers to alter the \emph{behavior} of malware to evade detection while still achieving their intended malicious effect~\cite{dynamic-analysis-survey,behavior-not-signatures}. Taking advantage of this insight, dynamic analysis relies on executing a sample in an isolated environment (sandbox) where the sample's behavior is observed and logged. The resulting features may include the spawning of processes, the creation or deletion of files, and the modification of registry keys (Figure~\ref{fig:mainfig}, panel 1). 

The machine learning community has been slower to adopt dynamic features as inputs to their malware detection models, primarily due to a lack of available datasets. A series of papers from Microsoft Research~\cite{microsoft1,microsoft2,mtnet} explored malware detection from sequences of API calls using logistic regression, RNNs, and deep feedforward networks, respectively, but the datasets used were proprietary, hindering academic efforts to reproduce and build on that research. Recognizing the need for more publicly available sandbox data, in 2022 Avast Software and Czech Technical University released the Avast-CTU dataset~\cite{avast-ctu} containing nearly 50,000 malware behavior reports; unfortunately, no reports were released for benign samples, restricting the applicability of that dataset. More recently, the Quo Vadis~\cite{quovadis} dataset, generated using the SpeakEasy Windows kernel emulator~\cite{speakeasy}, has facilitated research on a more expansive set of behavioral features, and newer work has found that incorporating file path information~\cite{quovadis}, network operations, and registry accesses~\cite{nebula} in addition to API call sequences can improve the performance of deep learning architectures. Despite these advances, malware detection from behavioral features using ML is far from solved.~\cite{quovadis} and~\cite{nebula} are limited by the vocabulary size and context windows of their underlying neural models, respectively, forcing them to drop uncommon API sequences and truncate behavior reports before providing the resulting incomplete inputs to their detection systems. Furthermore, in a detailed analysis of ML classifiers trained on behavioral features,~\cite{behavior-not-solved} found that many of these models were prone to learning spurious correlations instead of recognizing true indicators of maliciousness. As such, these models failed to generalize from their training sets to real-world test data.

\subsection{Concept Drift}
Concept drift refers to the phenomenon where the distribution of the target data changes over time, often between model training and deployment~\cite{concept-drift-background}. Concept drift is a pervasive problem in the area of malware detection due to the ongoing cat-and-mouse game between attackers and defenders in this space. As defenders learn of vulnerabilities and attacks, patch software, and update detection systems, adversaries adapt by finding new exploits or modifying existing ones to evade defenses. In fact,~\cite{needles-in-haystack} found that malware developers often submit early versions of their products to public detection tools (e.g. VirusTotal~\cite{virustotal}), review the outputs, and modify their code until they create a fully undetectable sample. Under this paradigm, machine learning classifiers trained at fixed points in time will gradually experience degradation in performance, as the malware they see at test time no longer resembles the samples they were originally taught to recognize~\cite{bodmas}.
One standard benchmark for concept drift in malware detection is the BODMAS dataset~\cite{bodmas}, which includes the date when each sample first appeared.

The predominant technique for countering concept drift is active learning~\cite{active-learning}, where, periodically, new samples are labeled by experts, added to the training set, and the classifier re-trained on this fresh data. The samples to label are often selected using techniques from out-of-distribution detection~\cite{ood-detection}. While active learning is effective (see~\cite{bodmas},~\cite{android-malware-continuous}), it is costly and time-consuming due to the need for expert labeling.

\section{Leveraging Behavior Reports with LLMs}
\subsection{The Behavior Report Dataset}
\label{sec:behavior-dataset}
We created our behavior report dataset by downloading a behavior report (collected in a sandbox) for each sample in the BODMAS Malware Dataset~\cite{bodmas} via VirusTotal~\cite{virustotal}. VirusTotal has six in-house sandboxes for the Windows operating system\footnote{https://docs.virustotal.com/docs/in-house-sandboxes} and integrates over a dozen third-party sandboxes;\footnote{https://docs.virustotal.com/docs/external-sandboxes} samples may contain behavior reports from multiple different sandboxes. We included in our dataset reports from the seven most frequent sandboxes: VirusTotal Cuckoofork, Tencent HABO, Zenbox, CAPE Sandbox, Rising MOVES, VirusTotal Jujubox, and Microsoft Sysinternals. We sanitize these reports to retain only information about the behavior of the sample, and remove all fields that are computed via static analysis: signature\_matches, mbc, mitre\_attack\_techniques, tags, verdicts, verdict\_confidence, sigma\_analysis\_results, ids\_alerts, verdict\_labels. For this reason, we also excluded reports from the CAPA ``sandbox'', as it is not a true sandbox, but a static analysis library. We did not otherwise normalize or truncate the behavior reports.

The original BODMAS dataset consists of 57,293 malicious and 77,142 benign Windows PE files. After querying sandbox reports from VirusTotal and applying the filters described above, we obtained behavior reports for 50,090 malware samples and 40,803 benign samples. Each behavior report is represented as a JSON; an example behavior report is displayed on the left-hand side of Figure~\ref{fig:mainfig}. As in BODMAS~\cite{bodmas}, we subdivided the dataset into 13 months (2019-09 to 2020-09) based on when a sample first appeared, with all samples appearing prior to 2019-09 being placed in the 2019-09 bucket. For all experiments, we trained on the 2019-09 data and tested on all subsequent months. To tune hyperparameters, we treated the 2019-10 data as a validation set.

\subsection{Producing Verified Malware Detection Rules}
\label{sec:rules}
We used Google's Gemini-3-Pro-Preview~\cite{gemini_3_pro} large language model with high reasoning to analyze the behavior reports. (Our full Gemini prompt can be found in Appendix~\ref{sec:appendix-verdict-prompt}.) The middle panel of Figure~\ref{fig:mainfig} shows an example of such analysis. As indicated by the yellow, blue, and pink highlights, the LLM is able to identify several notable behaviors from the sandbox report and reason about their likely malicious intent.

\begin{table}[t]
    \centering
    \begin{tabular}{c c c c}
        \toprule
        & \multicolumn{3}{c}{Predicted Label} \\ 
        True Label & \textbf{Malware} & \textbf{Benign} & \textbf{Error} \\ 
        \midrule
        \textbf{Malware} & 48.1\% & 1.6\% & 0.3\% \\ 
        \textbf{Benign} & 4.4\% & 45.6\% & 0\% \\ 
        \bottomrule
    \end{tabular}
    \caption{Confusion matrix for Gemini-3 malware detection from behavior reports.}
    \label{tab:llm-alone}
\end{table}

\begin{figure*}[t]
    \centering
    \includegraphics[width=\linewidth]{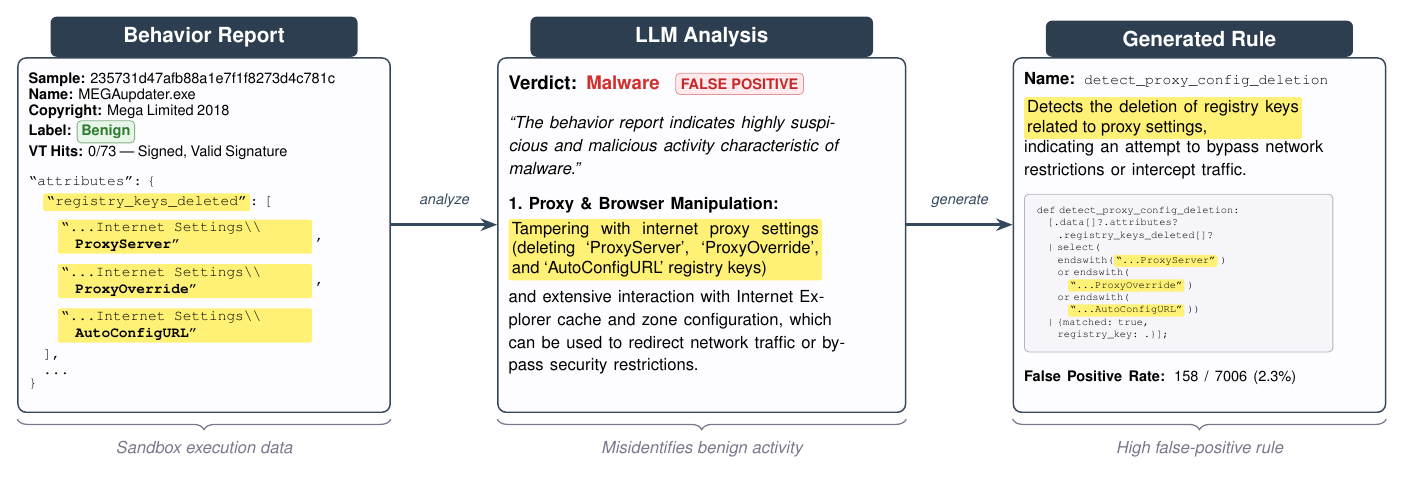}
    \caption{Rules can reduce LLM false positives. The LLM categorizes this sample as malware based on a suspicious behavior in the sandbox report. However, when this behavior is encapsulated in a rule and evaluated, it is proven to occur in small but meaningful percentage of benign samples. Thus, this behavior is not a reliable indicator of maliciousness.}
    \label{fig:case-study-llm-fp}
\end{figure*}

To measure the feasibility of using an LLM to classify samples as malware or benign based on their behavior report, we randomly selected and evaluated Gemini on 500 malicious and 500 benign behavior reports. The confusion matrix is shown in Table~\ref{tab:llm-alone}. Gemini's recall is very strong, correctly identifying 96.2\% of malware. (This number rises to 96.8\% if errors, which manual inspection reveals to be ``PROHIBITED\_CONTENT'' violations, are treated as malicious verdicts.) However, Gemini's false positive rate (FPR), at 8.8\%, is too high to make this method feasible for practical use, due to the overwhelming preponderance of benign files in real-world data feeds.

Figure~\ref{fig:case-study-llm-fp} shows how LLMs can be led astray when making isolated decisions on samples. The sample shown in the first panel is benign with confidence---it has zero hits on VirusTotal, is signed with a valid signature, and has a registered copyright. However, Gemini labels the file as malware, in part because the sample modifies internet proxy settings, which Gemini deems suspicious (second panel). However, though this behavior might be uncommon for benign executables, it is not inconsistent with benign intent. We asked Gemini to formalize this suspected malicious behavior into a detection rule (Figure~\ref{fig:case-study-llm-fp} third panel), and executed this detection rule against all benign samples that occur in the first month of our dataset. The rule flagged 158 / 7006 samples, giving the captured behavior a false positive rate of 2.3\%. 

This case study exemplifies a key benefit of rule-based detection systems---rules can be verified against a validation set of labeled data. Traditionally, malware detection rules have been hand-written by malware analysts, requiring significant time and expertise to develop. However, we found that large language models are able to automate the process of turning indicators of maliciousness into detection rules. Specifically, we prompted Gemini to identify the key behaviors that distinguish a given malware sample from benign software and, for each behavior, create a detection rule. The full prompt can be viewed in Appendix~\ref{sec:appendix-rule-gen-prompt}. We asked Gemini to write rules in the JQ language~\cite{jq_lang} because it is a standard filtering language for JSON files and we observed that Gemini was comfortable with its specifications.

We created 13,998 malware detection rules from the 4,307 malware samples in the first month of our behavior report dataset. We then relied on the 7,006 benign samples in our month-one data for filtering. Out of the initial tranche of rules, we kept rules that satisfied the following ``good rule'' criteria:
\begin{enumerate}
    \item Compiles successfully (has correct JQ syntax)
    \item Matches the sample it was generated from
    \item Has FPR = 0 on the month-one dataset
    \item Errors on fewer than 10\% of files in the  month-one dataset
    \item Finishes running within five minutes
\end{enumerate}
At the conclusion of this verification step, we were left with 8,648 ``good'' rules. An example of a good rule can be found in the rightmost panel of Figure~\ref{fig:mainfig}, while an example of a bad rule can be found in the rightmost panel of Figure~\ref{fig:case-study-llm-fp}.

A rule-based detection system suggests a natural confidence measure---the more rules that detect a sample, the more malicious behaviors it has, and the more confident we can be that that sample is malware. While this intuition is correct, the approach is complicated by the presence of duplicate rules in our rule-set. Since many malicious indicators are shared across malware samples, and we independently created rules from each sample, our initial rule-set contains multiple instances of rules detecting the same behavior. 

We de-duplicated the rules by clustering, then counted the number of rule clusters, rather than the number of individual rules, that hit each sample. We clustered rules by rule name, removing the ``rule\_'' prefix when present and removing any duplicated rule names. As features, we used character n-grams of size 3--5, with TF-IDF weighting, as well as token n-grams of size 1--3, tokenized on underscores and alphanumeric boundaries. We used the HDBScan clustering algorithm~\cite{hdbscan} with a minimum cluster size of 3, minimum sample size of 1, cosine distance metric, and excess of mass cluster selection method. Our choice of features and hyperparameters were informed by ablation experiments (Appendix~\ref{sec:appendix-clustering}).

\begin{figure}
    \centering
    \includegraphics[width=0.6\linewidth]{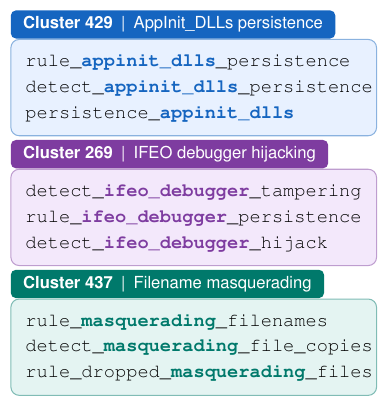}
    \caption{Rule clusters capture distinct behaviors.}
    \label{fig:clusters}
\end{figure}

From 8,648 good rules, we were left with 704 clusters and 1,692 singleton rules, which we preserved as clusters of size one. For non-singleton clusters, the median cluster size was 5, the mean cluster size was 9.88, and the maximum cluster size was 155. Figure~\ref{fig:clusters} shows a few clusters and the names of some of their constituent rules.

\subsection{Our Approach: Malware Detection with Behavior-Based Rules}
\label{sec:rules-results}
Having clustered the rules, we make classification decisions as follows:
\begin{enumerate}
    \item If no rules match a sample, label it as benign
    \item If more than $\tau$ rule cluster(s) match a sample, label it as malicious
    \item Otherwise, ask the LLM to label the sample based on the behavior report
\end{enumerate}
We consider a rule cluster to match a sample if ANY rule from that cluster matches the sample. If at most $\tau$ rule clusters match a sample, we consider the rules verdict on that sample to be uncertain, and defer to the LLM to classify it.

\begin{figure}[t]
    \centering
    \includegraphics[width=0.99\linewidth]{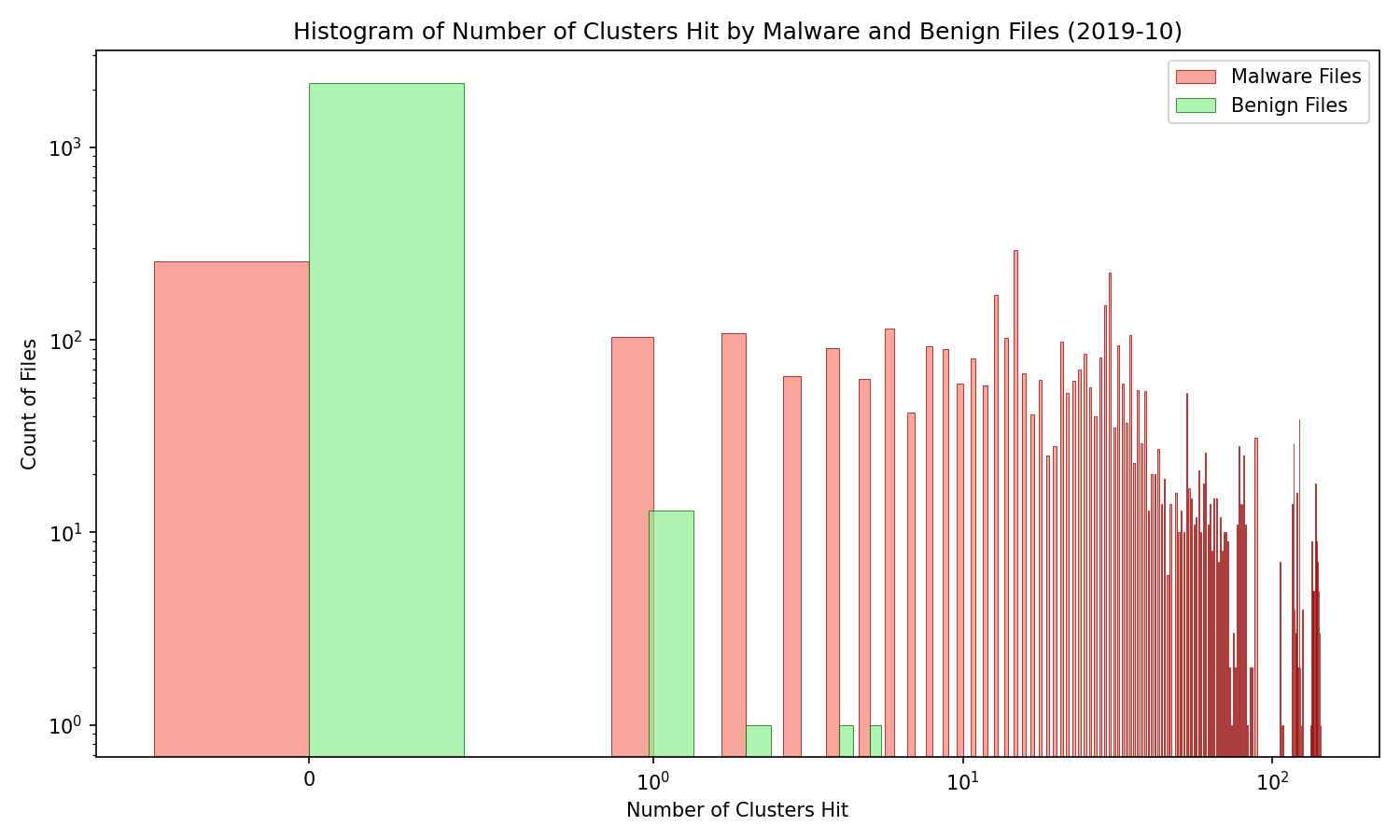}
    \caption{ Histogram of malicious and benign samples by number of rule clusters hit on the validation month (2019-10). Benign samples are less likely to be hit by multiple rule clusters.}
    \label{fig:cluster_distribution}
\end{figure}

The threshold $\tau$ is a hyperparameter that controls the tradeoff between false positives and false negatives.  A larger threshold reduces false positives and better tolerates imperfect rules (a single odd behavior might not necessarily mean the sample is malicious), but also risks missing some malware.
The threshold also affects the cost of classification: asking the LLM to label a sample incurs some cost, and the larger $\tau$ is, the more samples  need to be labeled with the LLM. As our chief concern is FPR, we chose to use $\tau=6$ for our primary experiments. Performance on our validation data (month two data; see Figure~\ref{fig:cluster_distribution}) indicated this threshold would give us the opportunity to correct nearly all false positives.

\begin{figure*}[htp]
    \centering
    \includegraphics[width=0.99\linewidth]{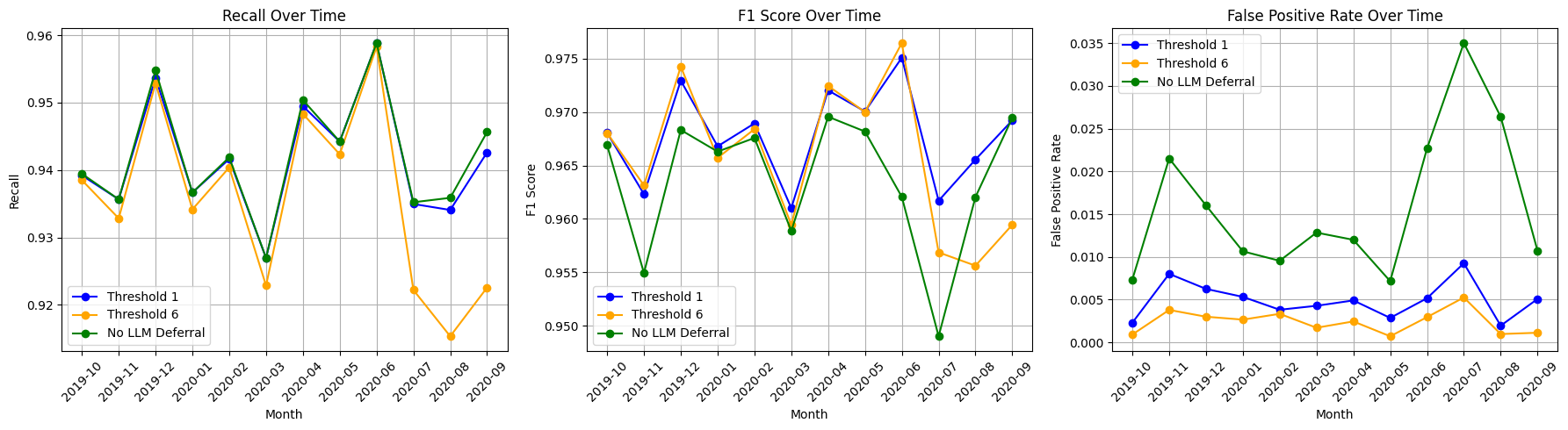}
    \caption{Performance of behavior-based malware detection rules at different LLM deferral thresholds.}
    \label{fig:rule_thresholds}
\end{figure*}

Figure~\ref{fig:rule_thresholds} highlights the impact of $\tau$ more broadly, showing the performance of our rule-based classification scheme at $\tau=0$ (no LLM deferral; only rules), $\tau=1$, and $\tau=6$. When using the rules alone ($\tau=0$), our false positive rate never exceeds 3.5\% in a single month, a 2.5x reduction over the 8.8\% FPR achieved when using the LLM alone (Table~\ref{tab:llm-alone}). With strategic deferral, the FPR is reduced to 1\% at threshold $\tau=1$ and 0.5\% at threshold $\tau=6$. Rule-based classification does come at a small cost to recall, which drops from 96\% with the LLM alone to 94\% with rules, except for the final few months at $\tau=6$, where the decrease is more pronounced. 

\begin{figure*}[htp]
    \centering
    \includegraphics[width=0.95\linewidth]{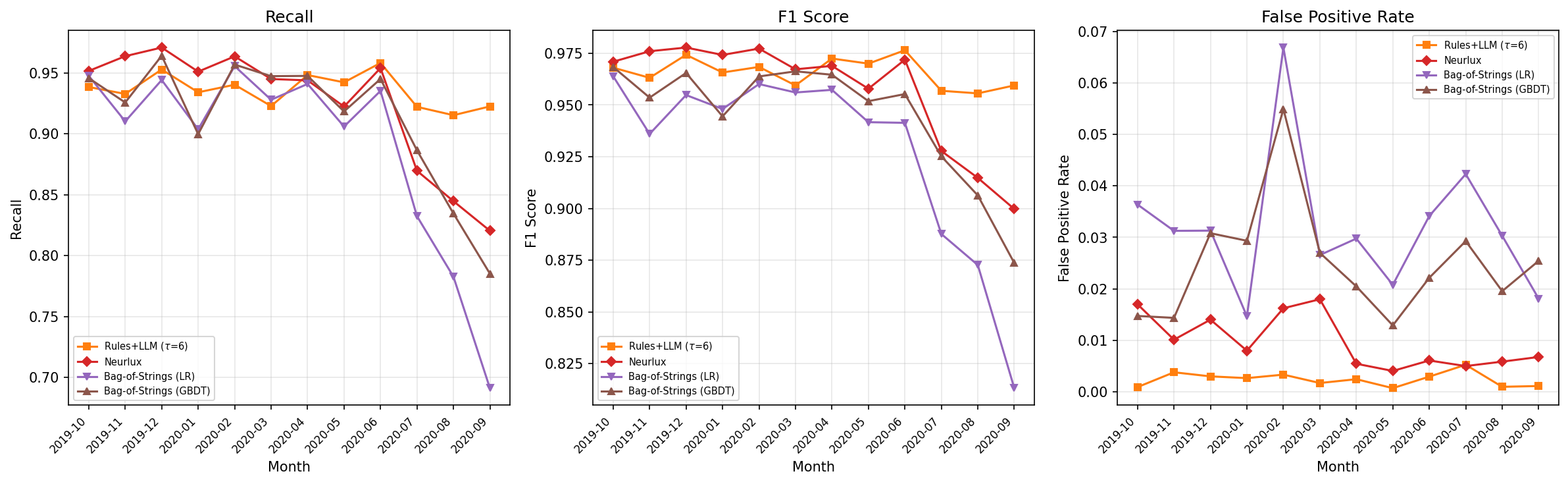}
    \caption{LLM-crafted malware detection rules outperform traditional ML methods applied to behavior reports.}
    \label{fig:dynamic-baselines}
\end{figure*}

\begin{figure*}[htp]
    \centering
    \includegraphics[width=0.95\linewidth]{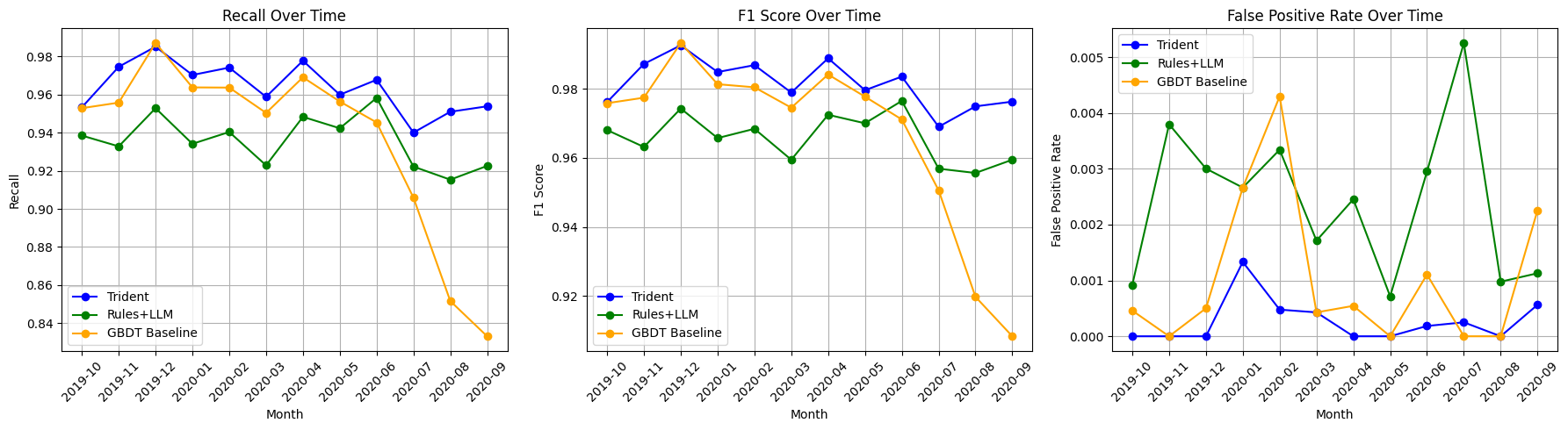}
    \caption{Performance over time of three different malware detection schemes: a gradient-boosted decision tree trained on EMBER features, behavior-based rules with an LLM deferral threshold of $\tau = 6$, and Trident, which combines a decision tree, rules, and LLM analysis under a majority vote.}
    \label{fig:trident_results}
\end{figure*}

We compare our behavior-based rules approach to other dynamic analysis baselines relying on behavior reports. Specifically, we compare to prior work Neurlux~\cite{neurlux}, and two simple bag-of-strings baselines. Our Neurlux implementation is derived from the code implementation released by its authors; we keep the same model architecture (CNN, BiLSTM, and Attention layers), hyperparameters, and  behavior report components (registry, file, network, dll, mutex, and api) as the original paper. 
For the bag-of-strings baselines, we convert the JSON behavior reports into key-value pairs (e.g. (``modules\_loaded'', ``kernel32.dll'') ), then use feature hashing to create feature vectors of size 4,096. Next, we train a logistic regression model and a gradient-boosted decision tree model over the feature vectors.
To match the setting of the rules-based experiments, we train all baseline models on the first month of data in our dataset (2019-09), validate on the second month of data when applicable, and evaluate on the subsequent months. 

Our baseline comparisons are shown in Figure~\ref{fig:dynamic-baselines}. Our rules outperform both bag-of-strings methods in F1-score (middle plot) in all-but-one month. Neurlux holds a slight advantage over the rules in F1-score in the first six months, but this trend reverses in the last six months, as both Neurlux and the bag-of-strings models suffer from concept drift, while the rules hold steady. Additionally, the rules have an appreciable advantage in false positive rate (rightmost plot); its trend line is consistently below, and much more stable, than that of other methods.

It is notable that our behavior-based rules are markedly more resilient to concept drift than the other behavior-centric approaches tested. We hypothesize that this is because our rules are generated by LLMs, trillion-parameter networks pre-trained on the entire internet, and thus are able incorporate common knowledge, reason over generalizable patterns, and avoid the types of spurious correlations that have been shown to confound other models. For example, while investigating the common failure modes of malware classification systems that do not transfer from one source of behavior reports to another,~\cite{behavior-not-solved} discovered that for one of the sandbox datasets they examined, ``Spotify'' was both prevalent (in 3\% of all samples) and predictive (occurring exclusively in malware). While traditional machine learning approaches would attempt to exploit this correlation to minimize training loss, large language models generating rules will be aware that ``Spotify'' is a popular music streaming service, not a reliable indicator of maliciousness.

\subsection{The Impact of Data Leakage} 
\label{sec:data-leakage}
One weakness of our experimental setup is that the publication date of the dataset we rely on, BODMAS~\cite{bodmas}, precedes the knowledge cutoff of the LLM we employ, Gemini-3-Pro-Preview~\cite{gemini_3_pro}, by several years. Therefore, we cannot be sure that the LLM has not seen, or worse, memorized, the samples we are evaluating on in its pretraining. Powerful reasoning models are necessary for the complex tasks we require of the LLM; no such models exist with knowledge cutoffs prior to our dataset's publication date. On the other hand, collecting, maintaining, and obtaining large datasets is a known challenge in malware research~\cite{raffmalwaremlchallenges}; we have only limited API access to popular services like VirusTotal~\cite{virustotal} and MalwareBazaar~\cite{malwarebazaar} from which to procure new data.

We evaluated whether our results might be impacted by data leakage by gathering a small malware dataset from 2026, after the LLM's knowledge cutoff.
Despite our resource constraints, we were able to collect behavior reports for nearly 5,000 PE malware samples first seen in 2026, more than a year past the knowledge cutoff for the Gemini 3 model series (January 2025). We downloaded a list of hashes from MalwareBazaar and used the family information available to perform stratified, round-robin sampling over the 520 malware families present; this maximized the diversity and representativeness of our sample. We next queried VirusTotal to secure behavior reports corresponding to these hashes, and filtered out reports from sandboxes not present in our BODMAS dataset (see Section~\ref{sec:behavior-dataset}). Finally, to best mirror our experimental setup with BODMAS, we assigned the 4,307 malware samples to the training split; thus, we are using an equivalent number of samples to generate the ruleset in each experiment. The remaining 589 malware samples were left for testing. We could not collect new benign samples as MalwareBazaar only hosts lists of malicious hashes---benign data often cannot be distributed due to copyright concerns~\cite{ember}.

We used Gemini-3.1-Pro-Preview to generate rules from the samples in our 2026 training corpus. (Access to Gemini-3-Pro-Preview was deprecated while this experiment was in progress, but Gemini-3 and Gemini-3.1 have the same knowledge cutoff.) We then filtered the rules using the same set of benign samples, and the same good rule criteria, as in the BODMAS experiments. After filtering, we used the surviving rules to measure recall on our 2026 test set. We labeled a sample as malware if it was hit by any rule, and benign otherwise; under this criterion, we measured a recall of 93.9\%, close to the 94.2\% recall of the behavioral rules alone on BODMAS (Table~\ref{tab:results-summary}). 
As the performances in our pre- and post-knowledge-cutoff experiments are comparable, we conclude that data leakage is not the reason for the success of our rules-based approach.

\section{Trident: Combining Static and Dynamic-Based Detection}
In this section, we demonstrate that malware detection from static features and malware detection from dynamic features are complementary approaches and introduce a new system, Trident, that combines methods from both paradigms to achieve superior results.
\subsection{Contrasting Static and Dynamic Detection Results}
\label{sec:static-vs-dynamic}
\noindent \paragraph{Baseline:} As a baseline, we compared our rule-based detection to the gradient-boosted decision tree (GBDT) classifier, trained on EMBER static features~\cite{ember}, introduced in the original BODMAS paper~\cite{bodmas}. We followed the same training, validation, and testing procedures described in~\cite{bodmas}, with the only difference being that we restrict the dataset to samples for which we have behavior reports, as described in Section~\ref{sec:behavior-dataset}. We trained the baseline model until it reached a FPR of 0.1\% on the validation set. 

We compared to a behavioral rule-based detection system, with rules generated by an LLM from behavior reports. We used a threshold of $\tau=6$ rule clusters, as discussed in Section~\ref{sec:rules-results}, as it is the nearest equivalent in terms of FPR. Results are shown in Figure~\ref{fig:trident_results}. For the first eight months of test data, the F1 score (middle panel) of the rules-based method (Rules+LLM) lags 1--2 percentage points behind the gradient boosted decision tree (GBDT); then, the GBDT's performance drops precipitously, consistent with the findings of the original BODMAS paper~\cite{bodmas}, while the Rules method remains fairly steady. These trends in F1 score are entirely driven by trends in recall (Figure~\ref{fig:trident_results}, left side). We investigated false negative samples from both methods to understand the sources of these gaps in recall.

\noindent \paragraph{Samples Missed by the Rules:} When we examined malware samples missed by the rules, we overwhelmingly found that their associated behavior reports were sparse, containing little-to-no data that could be leveraged by a detection rule. Detecting and evading sandboxes is a well-known aim of malware authors~\cite{sandbox-evasion}; when successful, such malware will not display its malicious behaviors when detonated in a sandbox. Additionally, we noticed several behavior reports from samples that crashed during sandbox execution among the false negatives; crashing prematurely would also limit the amount of malicious behavior data captured in a report. This reveals a fundamental limitation of relying on behavior reports: the rules cannot find what isn't present in the data. By contrast, static analysis is unaffected by sandbox evasion or crashing samples.

\begin{figure}
    \centering
    \includegraphics[width=0.95\linewidth]{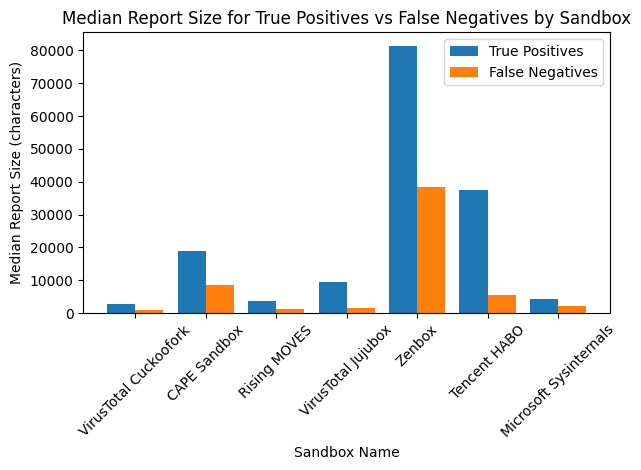}
    \caption{Median behavior report size for true positives and false negatives of the behavior-based detection rules, by sandbox. On every sandbox, false negatives are correlated with much less informative behavior reports.}
    \label{fig:fn_report_sizes}
\end{figure}

Figure~\ref{fig:fn_report_sizes} compares the median report size of true positive and false negative samples from each sandbox. Samples that are missed by the rules (false negatives) have much shorter behavior reports than detected samples (true positives), supporting the intuition that Rules-based detection fails primarily when the behavior report contains insufficient data from which to make a decision.

\begin{table}
    \centering
    \resizebox{0.48\textwidth}{!}{
    \begin{tabular}{l | c c c c | c | c}
        \toprule
        \multirow{2}{*}{\textbf{Month}} & 
        \multicolumn{2}{c}{\textbf{FNR Rules}} & 
        \multicolumn{2}{c|}{\textbf{FNR GBDT}} & 
        \textbf{$\Delta$ GBDT$-$Rules} &
        \textbf{Pct. New} \\ 
         & 
        \textbf{(Seen)} & 
        \textbf{(New)} & 
        \textbf{(Seen)} & 
        \textbf{(New)} & 
        \textbf{(New)} &
        \textbf{Fams.}\\
        \midrule
        10/19 & 5.42\% & 24.14\% & 3.64\% & 50.34\%  & 26.20\% & 3.41\%\\ 
        11/19 & 6.38\% & 8.00\% & 3.16\% & 46.67\%  & 38.67\% & 3.52\%\\ 
        12/19 & 4.40\% & 10.67\% & 1.65\% & 25.33\% & 14.66\% & 2.02\%\\ 
        01/20 & 6.08\% & 14.06\% & 3.76\% & 34.38\% & 20.32\% & 3.19\%\\ 
        02/20 & 5.04\% & 26.03\% & 3.61\% & 29.45\% & 3.42\% & 3.66\%\\ 
        03/20 & 6.68\% & 23.81\% & 4.16\% & 24.40\% & 0.59\% & 3.62\%\\ 
        04/20 & 4.87\% & 6.64\% & 2.92\% & 10.18\% & 3.54\% & 5.17\%\\ 
        05/20 & 4.68\% & 11.74\% & 3.41\% & 11.07\% & -0.67\% & 12.69\%\\ 
        06/20 & 3.94\% & 6.56\% & 4.81\% & 8.20\% & 1.64\% & 6.83\%\\ 
        07/20 & 6.88\% & 5.16\% & 11.13\% & 7.46\% & 2.30\% & 23.23\%\\ 
        08/20 & 5.67\% & 13.78\% & 14.43\% & 18.50\% & 4.72\% & 9.15\%\\ 
        09/20 & 3.43\% & 16.76\% & 15.32\% & 26.96\% & 10.20\% & 15.07\%\\ 
        \bottomrule
    \end{tabular}%
    }
    \caption{Comparison of False Negative Rates (FNR) across Rules (with LLM deferral threshold $\tau=6$) and gradient-boosted decision tree (GBDT) models for families present in (Seen) and not present in (New) the training data. Column 7 shows the percentage of malware samples each month from families not in the training data.}
    \label{tab:fnr_comparison}
\end{table}

\noindent \paragraph{Samples Missed by the Decision Tree:} We find that false negatives from the GBDT model are heavily driven by concept drift. There are two forms of concept drift in the malware detection problem space: existing families evolve gradually over time, as new variants emerge, and entirely new families appear as new exploits are discovered. Table~\ref{tab:fnr_comparison} shows the false negative rate (FNR) for the Rules+LLM method and the GBDT model on seen (in the training data) and unseen (not in the training data) families each month. Both techniques do better on seen families than unseen families (columns 2 vs 3 and 4 vs 5); however, in 11 out of 12 months, the GBDT FNR on new families is higher than the Rules+LLM FNR on new families (column 6), proving that detection from behavior-based rules generalizes better to unseen malware families than static analysis. This difference has a larger impact on recall as time goes on and the percentage of malware samples from new families (Table~\ref{tab:fnr_comparison} column 7) increases, a classic instance of concept drift, and eventually causes GBDT recall to drop below Rules+LLM recall (Figure~\ref{fig:trident_results} left panel). Table~\ref{tab:fnr_comparison} also reveals the effect of concept drift among seen families on the GBDT classifier; in the first nine months of the dataset, FNR among seen families (column 4) is consistently less than 5\%, but jumps to 15\% in the last two months of the dataset. By contrast, the FNR for the Rules+LLM method on seen families (column 2) remains in a narrow band between 3\% and 7\% for all months. 
This further magnifies the recall gap between the two methods in later months of data. 

\begin{figure*}
    \centering
    \includegraphics[width=0.95\linewidth]{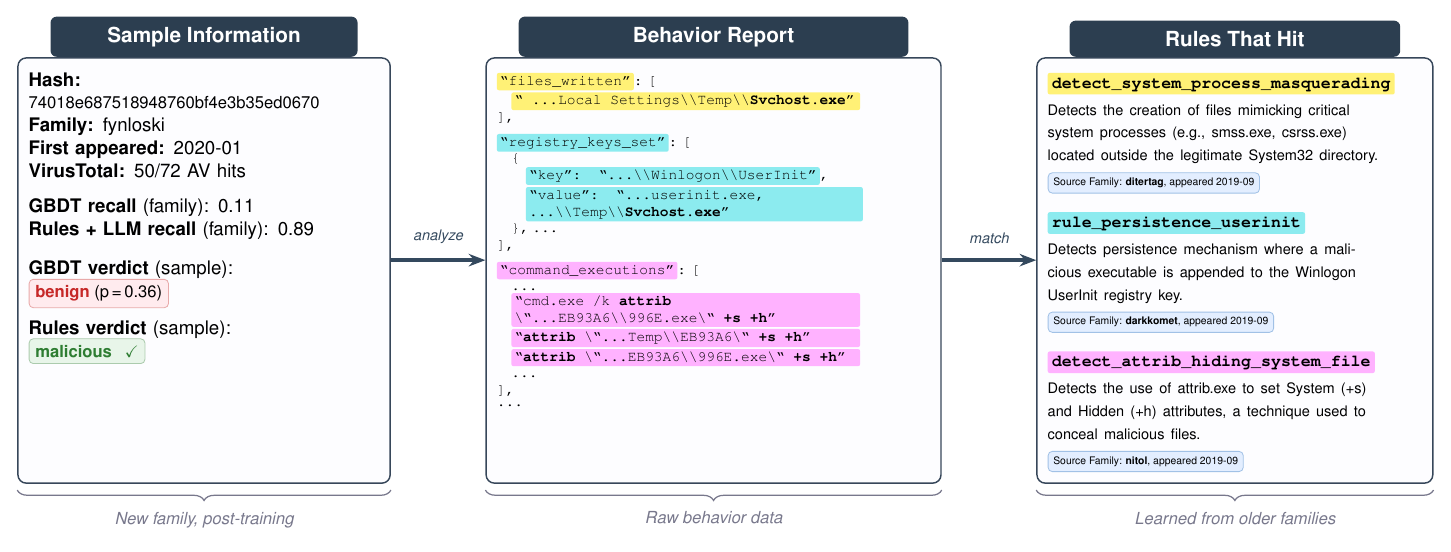}
    \caption{Behavior-based rules guard against concept drift. This sample, from a family not present in the training data, goes undetected by the gradient-boosted decision tree trained on static EMBER features. However, due to shared behavior with several older malware families, it is detected by multiple behavioral rules.}
    \label{fig:case-study-gbdt-fn}
\end{figure*}

We posit that behavior-based rules are more robust to concept drift than static-feature classifiers because while the static features of a file may be easily manipulated by attackers, core malicious techniques are slower to evolve. Figure~\ref{fig:case-study-gbdt-fn} shows an example of a malicious file that is missed by the GBDT classifier but detected by the rules. This sample is from the ``fynloski'' family, which first appears in 2020-01, four months after the classifier is trained. Likely due to the fact that it was not trained on any samples from the ``fynloski'' family, the GBDT classifier labels this sample as benign with a score of 0.36 (with 0.0 being certain benign, and 1.0 being certain malware). However, the rules correctly identify the sample as malware, flagging several indicators of maliciousness in the behavior report. These behaviors are shared with malware samples in the ``ditertag'', ``darkkomet'', and ``nitol'' families, respectively, which all appear in the first (training) month of the dataset.

The behavioral features identified by our rules stand in stark contrast to the static EMBER features employed by the GBDT classifier. EMBER features include attributes like file size, imported and exported functions, the size and entropy of each section, byte histograms, a histogram of printable characters within strings, etc.~\cite{ember}. These features are brittle: an adversary could inadvertently modify many of these characteristics while making routine updates to their malware, and a skilled attacker could deliberately manipulate such features by, for example, adding dummy code to the executable. It would be significantly more challenging to come up with a new persistence mechanism, or a new way of concealing malicious files, to alter the behavioral features detected by the rules in Figure~\ref{fig:case-study-gbdt-fn}, which helps to explain why these features are more robust to concept drift.

\subsection{The Best of Both Worlds}
\label{sec:trident-results}

\begin{algorithm}
\caption{Tiebreak algorithm for Trident majority-vote scheme}
\label{alg:tiebreak}
\algrenewcommand\algorithmicrequire{\textbf{Input:}}
\algrenewcommand\algorithmicensure{\textbf{Returns:}}
\begin{algorithmic}[1]
\Require $\textit{llm\_verdict} \in \{\text{`benign'}, \text{`malicious'}\}$, $\textit{gbdt\_prob} \in [0, 1]$
\Ensure Final classification label

\If{$\textit{llm\_verdict} == \text{`benign'}$}
    \State \Return \text{`malicious'} if {$\textit{gbdt\_prob} \geq 0.99$}, otherwise \text{`benign'}
\ElsIf{$\textit{llm\_verdict} == \text{`malicious'}$}
    \State \Return \text{`malicious'} if {$\textit{gbdt\_prob} \geq 0.5$}, otherwise \text{`benign'}
\EndIf
\end{algorithmic}
\end{algorithm}

As the discussion in Section~\ref{sec:static-vs-dynamic} indicates, static and dynamic-based detection methods have complementary strengths. Dynamic features change more slowly over time, making these approaches more robust to concept drift, where static analysis struggles. However, static analysis is needed when limited behavior information is available, due to malware evading sandbox analysis or crashing upon execution. To get the best of both worlds, we introduce Trident, which takes three inputs---a verdict (``malicious'' or ``benign'') from the GBDT classifier, a verdict (``malicious'', ``uncertain'', or ``benign'') from the behavior-based rules, and a verdict (``malicious'' or ``benign'') from the LLM (Gemini)---and conducts a majority vote. The GBDT verdicts are computed using the decision boundary calibrated at training time (0.983); for the rules, a ``benign'' label is assigned if no rules match, a ``malicious'' label is assigned if more than six rule clusters match, and an ``uncertain'' label is assigned otherwise. In the event of a tie (the rules verdict is ``uncertain'', the LLM and the GBDT model disagree), we use Algorithm~\ref{alg:tiebreak} to assign a verdict. If there is a tie and the LLM predicts benign, we assign a verdict of malicious only if the GBDT model is highly confident (predicted probability >= 0.99); if the LLM predicts malicious, we assign a verdict of malicious at a lower GBDT probability threshold (predicted probability >= 0.5). Only 1.20\% of classification decisions are made by the tiebreak algorithm. As an additional cost optimization, we only query the LLM for a verdict when the GBDT and Rules methods disagree.

The results are shown in Figure~\ref{fig:trident_results}. Trident's F1 score equals or surpasses that of GBDT alone and Rules+LLM alone, in every month, driven by equivalent gains in recall. Performance is also significantly more stable, highlighting improved resilience to concept drift. Notably, Trident's false positive rate is also lower than that of its components in 11 out of 12 months, and less than 0.1\% in 11 out of 12 months.

In addition to better performance, Trident offers large gains in interpretability over a decision tree model. Every ``malicious'' label is supported by either a set of rules that flagged the sample or an LLM explanation of key behavioral indicators of maliciousness, which analysts can easily review if they doubt a label's veracity. Though ``benign'' labels are traditionally harder to justify, as they primarily reflect the absence of ``malicious'' characteristics rather than the affirmative presence of ``benign'' characteristics, on some decision paths Trident also provides LLM explanations for ``benign'' labels that are available for analysts to review.
\paragraph{Comparison to Active Learning}
\begin{figure*}
    \centering
    \includegraphics[width=0.95\linewidth]{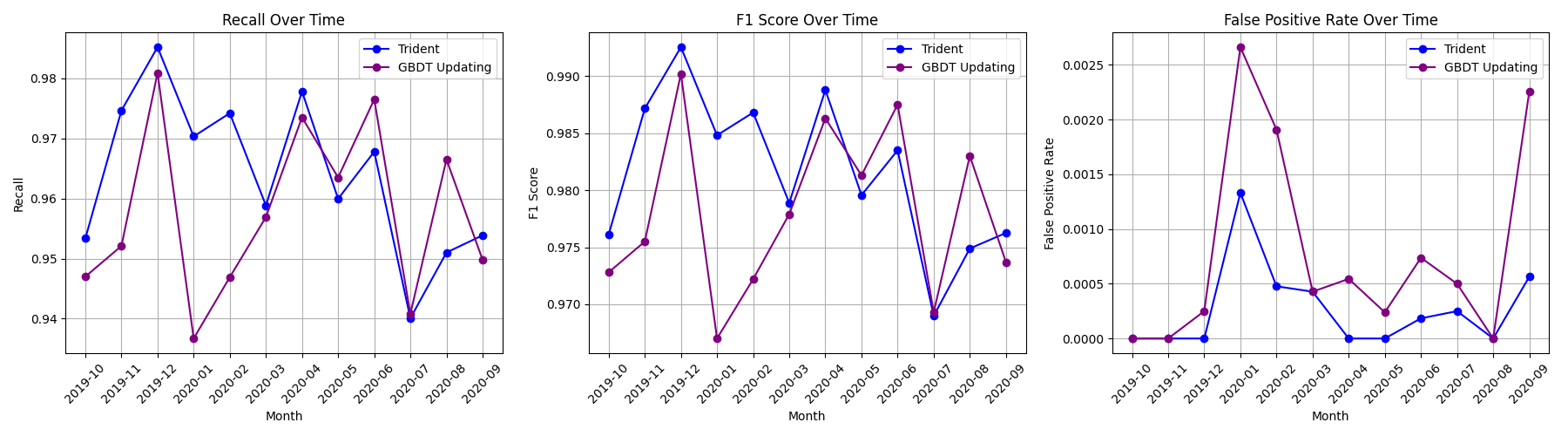}
    \caption{Comparison of Trident to a gradient-boosted decision tree (GBDT) with active learning. Each month, the GBDT model is retrained with an additional 1\% of labeled data from the previous month. Trident uses no new labels.}
    \label{fig:active-learning-comp}
\end{figure*}

One popular method for combating concept drift in malware classifiers is active learning, where, periodically, new labeled samples are added to the dataset and the classifier is retrained. In BODMAS~\cite{bodmas}, the authors show that by retraining a decision tree classifier each month with an additional 1\% of expert-labeled data from the previous month, they are able to greatly improve robustness to concept drift. Further, they show that randomly selecting which samples to retrain on has similar performance to more complex selection methods. In Figure~\ref{fig:active-learning-comp}, we modify our decision tree baseline from Section~\ref{sec:static-vs-dynamic} to the active learning paradigm, retraining every month with an additional, randomly sampled 1\% of ground-truth data from the prior month, and compare to Trident. (For our Trident GBDT verdicts, we use the original GBDT classifier, not the retrained one.) Trident is strictly better in terms of false positive rate, and competitive in terms of recall, without incurring the (high) cost of acquiring fresh labels and retraining a classifier each month. This shows that Trident can achieve resilience to concept drift at a far lower cost than prior methods, because it does not require human experts to label samples every month.

\section{Discussion}
We summarize our key findings, acknowledge limitations, and suggest directions for future work.
\subsection{Key Findings}
\begin{enumerate}
\item \textbf{LLMs can analyze malware behavior reports.} In Section~\ref{sec:rules}, we show that reasoning models can process semi-structured sandbox behavior reports, identify indicators of maliciousness, and render malicious/benign verdicts. They can also write high-quality detection rules that capture these indicators.

\item \textbf{We can minimize false positives by validating malicious indicators and strategically deferring to an LLM.} LLM false positive rates are too high for practical applications (Table~\ref{tab:llm-alone}). By having the LLM create detection rules based on suspected malicious indicators, verifying these rules on a validation set to ensure a low false positive rate, and only using an LLM verdict when the rules do not confidently predict a label, we achieve a $17.6\times$ reduction in the false positive rate (Section~\ref{sec:rules-results}).

\item \textbf{Behavior-based detection rules are more robust to concept drift than classifiers built on static features.} In Section~\ref{sec:static-vs-dynamic}, we show that our detection rules exhibit much more stable recall over time than a traditional decision tree model trained on static features (Figure~\ref{fig:trident_results}), and that this stability is due to better performance on new families, i.e. families that do not appear in the training set but emerge later in the test data (Table~\ref{tab:fnr_comparison}). 

\item \textbf{Static features complement behavior. Performance is maximized by Trident, which combines static-based methods, behavioral rule output, and LLM analysis via majority vote.} Rules-based detection struggles when behavioral information is sparse, which may occur when a sample evades sandbox analysis or crashes during execution (Section~\ref{sec:static-vs-dynamic}). This suggests an important role for static analysis -- filling in these gaps. Affirming this insight, in Section~\ref{sec:trident-results} we show that the best performance is achieved by Trident, a simple system that takes a majority vote over (i) the verdict from a decision tree trained on static features, (ii) the output of our behavioral detection rules, and (iii) the LLM analysis of the behavior report.

\item \textbf{Trident is as good as retraining a classifier monthly in protecting against concept drift.} Machine learning classifiers traditionally protect against concept drift through active learning, which entails regularly acquiring new labeled data and retraining the classifier. Trident achieves the same protection that active learning offers against concept drift simply by incorporating behavioral features during the initial month of system calibration (Figure~\ref{fig:active-learning-comp}), without the need for expensive relabeling or retraining, thereby reducing the cost of upkeep.

\item \textbf{Trident is more interpretable than traditional static feature classifiers without sacrificing performance.} Every ``malicious'' label offered by Trident is backed by either an LLM analysis of the sample, or rules that match the sample, which are simple for analysts to review. Affirmative explanations (LLM analyses) are even available for some benign samples, which is rare in malware detection, as a benign label frequently reflects the absence of malicious indicators rather than the presence of benign indicators, hindering explainability.
\end{enumerate}

\subsection{Limitations}
One limitation of our method is that it requires a sandbox behavior report for every sample. Running samples through a sandbox takes time, sometimes five or more minutes per sample~\cite{daily-malware}, so a production deployment would likely combine Trident with other mechanisms to handle easy-to-classify samples.

As discussed in Section~\ref{sec:data-leakage}, our study is also limited by the data available to us. While evaluating Trident on newer and larger datasets would allow us to strengthen our claims, we face budget, API, and copyright (for benign samples) restrictions that prevent us from collecting such datasets ourselves. Such challenges are pervasive in malware research, and it is not uncommon for researchers to rely on datasets that are even older than BODMAS~\cite{ember2024,beacon}. For a more comprehensive discussion of the difficulties in data collection for malware research, we refer to~\cite{raffmalwaremlchallenges}.

\subsection{Future Work}
\paragraph{Combining Static and Dynamic Detection Methods}
When combining static and dynamic-based detections, Trident employs a majority vote to determine the final label, which may overlook subtle interdependencies between classifiers. Recent work suggest that this limitation could be partially mitigated by using a stacked ensemble framework to integrate multiple models~\cite{ensemble-learning-1,ensemble-learning-android}. Training a meta-classifier model to weight classification signals based on historical accuracy may enhance Trident's synthesis of static and dynamic-based detection. 

\paragraph{Improving Rules}
Our current approach filters out rules that do not match our quality criteria, which leads us to discard roughly 38\% of rules. Future work could explore repairing these rules instead of discarding them entirely. One approach might be feeding the rejected rules back to the LLM for repair, with guidance as to how they originally fell short. Repairing bad rules has the potential to increase malware coverage and, in turn, reduce the incidence of ``uncertain'' rule verdicts at test time.

In addition to reducing rule false negatives, there is room to further reduce rule false positives. While investigating such false positives, we observed that our choice to use the JQ format for rules at times constrained rule quality. For instance, we noticed several rules that attempt to detect random strings (filenames, process names, etc.) as an indicator of malicious behavior. Unfortunately, with JQ, rules are confined to using regex filters to detect randomness; because regex operates on structural patterns rather than statistical properties, these filters are often more permissive than intended (e.g., \verb@\[a-z]{5,10}\.(exe|dll)@). In such cases, a lightweight postprocessing tool like pygarble~\cite{pygarble} can easily filter truly random matches (``qjfkhj") from non-random ones (``olepro"), helping to reduce false positives. We do not apply any postprocessing in our current approach; however, future work may consider allowing selective postprocessing to improve rule fidelity, or moving away from the JQ format to allow for more sophisticated rule construction.

\section{Related Work}
There is significant literature on using machine learning to do malware detection from static features. EMBER~\cite{ember} developed a set of hand-engineered features for PE files that have been widely adopted. By contrast, MalConv~\cite{malconv} showed that it was possible to do malware detection from raw bytes, without the need for feature engineering. Bhardwaj et. al.~\cite{bhardwaj} rely on static features to achieve a 99\% F1 score on BODMAS, the same dataset we use, but do not do a temporal split of their train and test data, whereas our work focuses on mitigating temporal concept drift. We refer readers to~\cite{survey-ml-malware} for a comprehensive overview of traditional machine learning techniques for malware detection.

Prior to the emergence of LLMs, MalBert~\cite{malbert} showed that transformers could be used to classify Android applications into specific malware categories. Since then, a body of work has been done leveraging LLMs for static analysis in malware detection. LLM-MalDetect~\cite{maldetect} uses a fine-tuned Mistral model trained on text features from Android APKs to classify samples, and TraceRAG~\cite{TraceRAG} uses snippets from decompiled source code to generate a human readable report in addition to classification; MalCVE~\cite{malcve} also generates an LLM summary from decompiled source code, then uses RAG to associate samples with known CVEs. 
AppPoet~\cite{AppPoet} uses an LLM to generate descriptions of the static features of an application, which are used by a DNN to classify samples, and produces a human-readable report. Unlike~\cite{AppPoet}, which feeds the features into a DNN for classification, LAMD~\cite{LAMD} directly uses the LLM as a classifier. By using an LLM to reason through an application's structure and provide a verdict as well as a report, it trades consistency and cost for reasoning capability. 

Initial research applying machine learning to behavioral features for malware detection focused on API call sequences~\cite{microsoft1, microsoft2,mtnet,zhang2020}. More recent work has expanded the feature set to include file paths~\cite{quovadis}, network operations, and registry accesses~\cite{nebula}, which has improved results. Richer features have not come for free, though, and these systems, struggling with large vocabulary sizes and input lengths, are forced to drop uncommon API sequences, normalize extensively, or truncate behavior reports to abide by the restrictions of older neural models. In contrast, the large language models that we employ for Trident are able to accommodate the full descriptiveness of our input data without any preprocessing. 

Some earlier work specifically applies methods from natural language processing to behavior reports for malware classification. MalDy~\cite{maldy} uses a traditional bag-of-words approach, while Neurlux~\cite{neurlux} employs a combination of CNNs, BiLSTMs, and attention networks to do malicious/benign classification. Our method differs from these precursors in three significant ways. First, on most samples, we don’t directly utilize the neural model to make a classification decision. Instead, our primary use for LLMs is to write malware detection rules; these rules are then deployed to make classification decisions at test time. Secondly, we use the newer, larger, and much more performant LLMs, as compared to the smaller, more specialized models used in these systems. Finally, we do not train a custom model for malware detection, but instead use off-the-shelf, generalist models. These models receive information about malware and malware analysis in their training data, but also have a broad range of additional expertises and capabilities, which they can draw on to supplement their malware detection proficiency. 

Two more recent works also bring together LLMs and behavior reports. MaLAware~\cite{malaware} focuses on explainability rather than classification, generating human readable reports by analyzing sandbox behavior. BEACON~\cite{beacon} uses LLMs to extract embeddings from behavior reports, which are then fed into a CNN for malware family classification. Our approach differs from that of~\cite{beacon} as we generate detection rules directly from behavior reports using LLMs, and perform benign-malicious classification, not malware family classification.

Alongside standard classification tasks, the problem of concept drift in malware detection has been studied extensively~\cite{tracking-concept-drift}. Transcend~\cite{transcend} focuses on timely identification of stale malware classifiers,~\cite{zero-days} measures the efficacy of deep-learning approaches at detecting zero-day samples, and CADE~\cite{CADE} identifies drifting samples.~\cite{bodmas} and~\cite{concept-drift-windows} explore remediation of drifting Windows PE malware classifiers through periodic retraining;~\cite{android-malware-continuous} takes a similar approach to Android malware classification.~\cite{concept-drift-RL} also tackles concept drift in Android malware classification, but uses techniques from deep reinforcement learning.

Our approach uses LLMs to generate behavior-based malware detection rules in an automated fashion. In the static analysis domain, AutoYara~\cite{autoyara} uses n-gram mining and biclustering to automatically generate YARA rules from sample files. Relatedly,~\cite{automl-for-malware} explores automated machine learning techniques for neural architecture search and hyperparameter optimization in deep learning-based static malware classifiers. Trident is also not the first to work to combine static and dynamic analysis.~\cite{quovadis} likewise synthesizes the results of static and behavioral-feature classifiers, but undergoes the added complexity of training a meta-model, rather than taking a majority vote.~\cite{integrated-static-dynamic} demonstrates improvements by marrying static and dynamic analysis at the feature level, training a single classifier on a mixture of feature types.

\section{Conclusion}
In this paper, we showed that machine learning-based malware detection can be improved by using LLMs and behavioral features. We demonstrated that LLMs can be used to analyze malware behavior reports and create interpretable, behavior-based malware detection rules at scale. These behavioral rules can be validated to facilitate low false-positive rates; even after low-precision rules are removed, we found that behavior-based detection was more robust to concept drift than traditional ML classifiers built on static analysis features. Static features can still offer valuable insight, especially when behavioral data is scarce, prompting us to design Trident, a malware detection system that combines a static-based ML classifier, behavioral rule output, and LLM analysis through a majority vote. Trident outperforms each of its component methods (static feature detection, behavior-based rules, LLM analysis) alone, and offers equivalent protection against concept drift as retraining a static-feature classifier monthly. In short, Trident provides superior performance and interpretable results without the cost of frequent retraining. 
\section*{Acknowledgments}

This material is based upon work supported by the Air Force Office of Scientific Research under award number FA9550-23-F-0014 in the amount of \$140,900, in addition to generous gifts from the Noyce Foundation, Google, and OpenAI. The authors would also like to thank Mona Wang, incoming Assistant Professor affiliated with The Citizen Lab at the University of Toronto, for her support.

\appendix

\bibliographystyle{plainurl}
\bibliography{references}
\cleardoublepage

\section{LLM Prompts}
\subsection{Verdict Prompt}
\label{sec:appendix-verdict-prompt}
You are an expert security analyst. Given the behavior report below, determine if the given sample is malicious or benign. Give a verdict of True for malicious or False for benign. Provide a detailed explanation for your conclusion. \\

\noindent Behavior Report: \{report\}
\subsection{Rule Generation Prompt}
\label{sec:appendix-rule-gen-prompt}
You are an expert security analyst. Given the VirusTotal behavior report below, identify the key behaviors, if any, that distinguish this malware sample from benign software. For each key behavior identified, create a detection rule using jq-style filters. The rule should be general enough to catch this malware sample and similar variants, but specific enough to avoid false positives on benign software. If the rule matches on a test file, it should return an array of results; otherwise, it should return an empty array. \\

\noindent Provide the rules in JSON format. If no distinguishing behaviors are found, return an empty list of rules. \\

\noindent Here is a sample rule and its JSON representation.

{\small
\begin{verbatim}
```
#Rule 1: Detect C2 Server Communication (specific IP)
def rule_c2_known_ip:
  [.data[]?.attributes?.ip_traffic[]?
  | select(.destination_ip == "88.198.101.58")
  | {{matched: true, ip: .destination_ip, 
      port: .destination_port, 
      protocol: .transport_layer_protocol}}
  ];
```
{{
  "rules": [
    {{
    "name": "rule_c2_known_ip",
    "description": "Detect C2 Server Communication",
    "rule":  
       "def rule_c2_known_ip:
          [.data[]?.attributes?.ip_traffic[]?
          | select(.destination_ip=="88.198.101.58")
          | {{matched: true, ip: .destination_ip, 
                port: .destination_port, 
                protocol: .transport_layer_protocol
             }}
          ];",
     }}
   ]
}}
\end{verbatim}
}

\noindent \\ Here is the behavior report to analyze: \{report\}

\section{Rule Clustering Ablations}
\label{sec:appendix-clustering}

\begin{table*}
\centering
\caption{Ablation study on deduplication parameters. We report the mean and standard deviation across runs. The \textbf{best} value for each metric is bolded, and the baseline configuration (name, N-Gram, 4, 2) is \underline{underlined}.}
\label{tab:clustering-ablation}
\begin{tabular}{llcc ccc}
\toprule
Input & Feature & Min Cluster Size & Min Samples & Recall & FPR & F1 \\
\midrule
description & N-Gram & 4 & 2 & $.9319 \pm .0008$ & $.0051 \pm .0000$ & $.9635 \pm .0004$ \\
rule & N-Gram & 4 & 2 & $.9328 \pm .0016$ & $.0064 \pm .0000$ & $.9636 \pm .0009$ \\
name & Jina-v3 & 4 & 2 & $.9327 \pm .0002$ & $.0057 \pm .0003$ & $.9637 \pm .0001$ \\
description & Jina-v3 & 4 & 2 & $.9328 \pm .0009$ & $.0047 \pm .0002$ & $.9640 \pm .0005$ \\
rule & Jina-v3 & 4 & 2 & $\textbf{.9375} \pm .0003$ & $.0064 \pm .0000$ & $\textbf{.9661} \pm .0002$ \\
\midrule
name & N-Gram & 2 & 2 & $.9311 \pm .0016$ & $.0042 \pm .0004$ & $.9632 \pm .0009$ \\
name & N-Gram & 3 & 2 & $.9295 \pm .0013$ & $.0041 \pm .0000$ & $.9624 \pm .0007$ \\
name & N-Gram & 5 & 2 & $.9307 \pm .0004$ & $.0048 \pm .0008$ & $.9629 \pm .0003$ \\
name & N-Gram & 6 & 2 & $.9315 \pm .0008$ & $.0055 \pm .0000$ & $.9631 \pm .0004$ \\
name & N-Gram & 8 & 2 & $.3181 \pm .0053$ & $\textbf{.0005} \pm .0000$ & $.4826 \pm .0060$ \\
name & N-Gram & 10 & 2 & $.3181 \pm .0053$ & $.0005 \pm .0000$ & $.4826 \pm .0060$ \\
\midrule
name & N-Gram & 4 & 3 & $.9319 \pm .0007$ & $.0044 \pm .0006$ & $.9636 \pm .0005$ \\
name & N-Gram & 4 & 4 & $.3746 \pm .0040$ & $.0005 \pm .0000$ & $.5449 \pm .0043$ \\
name & N-Gram & 4 & 5 & $.8690 \pm .0001$ & $.0014 \pm .0000$ & $.9296 \pm .0001$ \\
name & N-Gram & 4 & 6 & $.9027 \pm .0000$ & $.0018 \pm .0000$ & $.9484 \pm .0000$ \\
name & N-Gram & 4 & 8 & $.8191 \pm .0001$ & $.0018 \pm .0000$ & $.9001 \pm .0001$ \\
name & N-Gram & 4 & 10 & $.8832 \pm .0001$ & $.0032 \pm .0000$ & $.9372 \pm .0001$ \\
\midrule
name & N-Gram & 4 & 2 & $\underline{.9296} \pm .0013$ & $\underline{.0043} \pm .0004$ & $\underline{.9624} \pm .0006$ \\
\bottomrule
\end{tabular}
\end{table*}

We performed ablations on our rule clustering and deduplication setup. Specifically, we evaluated four hyperparameters: 
\begin{enumerate}
    \item the input to the HDBScan clustering algorithm -- either the rule name, rule description (a one-to-two sentence summary generated by the LLM at the time of rule creation), or rule text,
    \item the feature type (N-grams or output from the Jina v3 text embeddings model~\cite{jina-embeddings}),
    \item the minimum cluster size
    \item the minimum samples, which corresponds to the parameter $k$ used  when HDBScan calculates distance to a $k^{th}$ nearest neighbor. Larger values label more points as noise.
\end{enumerate}
As we are clustering as a form of deduplication, we measure effectiveness by randomly sampling one rule from each cluster and using this subset of rules to compute recall, false positive rate, and F1-score. We repeat the random sample five times, and report the means and standard deviations for each metric. In an effective clustering, where all rules in a cluster are semantic duplicates, a subset of rules selected this way will retain high recall and low standard deviation across multiple samples. We used the second month in our dataset (2019-10) for these validation purposes, and present the results of our ablation experiments in Table~\ref{tab:clustering-ablation}. The clustering method chosen for the main experiments in our paper -- based on rule name, with N-gram features, minimum cluster size of four, and minimum samples of two -- is displayed in the bottom row for ease of comparison.

For all combinations of input and feature types, and most combinations of the other hyperparameters, we observe that it is possible to achieve marginal improvements or declines in F1-score, but any improvements were so minor that we did not feel it necessary to change from our original approach (bottom row). We do find that some ranges of hyperparameters (minimum cluster size greater than 6, minimum samples greater than 3) were clearly detrimental to performance.

\end{document}